# Design Principles for Architectures of Technical Smart Service Systems

Nikola Pascher[1] and Jochen Wulf[1]

[1] ZHAW School of Engineering
Institute of Data Analysis and Process Design (IDP)
Technikumstrasse 81, CH-8401 Winterthur
nikola.pascher@zhaw.ch

**Abstract.** Successful smart services require seamless integration into existing corporate systems and an interdisciplinary approach that aligns the development of both business models and technical architectures. Multi-disciplinarity and co-creating with customers add a layer of complexity but are essential collaboration-schemes for validating the value proposition of smart services and building long-term customer loyalty. This paper explores these challenges and distills the design principles for the architectures of technical smart service systems, based on empirical data from architecture projects in two manufacturing companies. These principles contribute to the sparse academic literature on this topic and help practitioners navigate several design trade-offs commonly arising in smart service projects.

**Keywords:** Smart services, architecture, IIoT, design principles

## 1 Introduction

In the evolving landscape of the digital economy, industrial companies are standing at the brink of a transformative opportunity presented by the Industrial Internet of Things (IIoT) and Industry 4.0. These technological paradigms offer unparalleled potential for revolutionizing traditional manufacturing processes through the implementation of smart, data-driven services, subsequently leading to the creation of new value streams and the realization of significant cost efficiencies. The integration of smart services into business operations is increasingly recognized as a pivotal step in the innovation management strategies of traditional manufacturing firms (Beverungen et al., 2019). This recognition stems from the growing need to enhance operational efficiency, improve product quality, and offer custom, value-added services to meet changing consumer expectations.

However, the transition to adopting smart services is fraught with complex challenges. For these services to be viable, they must be sellable, scalable, serviceable, and maintainable (Götz et al., 2018). This necessitates the robust integration into existing corporate systems and mandates a multidisciplinary approach that harmonizes the development of both business models and technical architectures. Moreover, another dimension of complexity is introduced by the imperative to co-create with customers, which is critical for substantiating the value proposition of smart services and fostering customer loyalty.

Many firms, however, encounter substantial impediments when it comes to designing sustainable technical architectures that are pivotal for the successful deployment and scaling of smart service systems. The challenges often stem from a lack of comprehensive design principles that cater to the nuanced demands of these systems, encompassing their technical, economic, and sociotechnical aspects (Wolf et al., 2020). Consequently, there is a considerable gap in academic and practical knowledge regarding effective frameworks and methodologies for architecting technical smart service systems that are resilient, adaptable, and aligned with overarching business strategies.

To address this gap, we present a design science research project that introduces design principles for architectures of technical smart service systems. The paper is structured as follows. After discussion IoT platforms and architectural requirements of smart services in the related work section (section 2), we present our research methodology in section 3. In the results section (section 4), we describe our method for constructing and validating the proposed design principles. We present contributions to theory and practice in section 5 and conclude with a discussion of limitations and future research.

## 2 Related Work

### 2.1 IoT Platforms

The term "Internet of Things" (IoT) was first introduced by Kevin Ashton in 1999. It refers to a data network that enables automatic data gathering from physical objects using a unique identifier and, for example, radio frequency transmission (Ashton, 2009). An IoT platform is understood as a combination of hardware and software systems that execute the standard and universal functions of diverse IoT applications (Mineraud et al., 2016).

There is vast research on the anatomy of IoT platforms (Čolaković & Hadžialić, 2018). In an extensive literature research, Barros et al. (2022) distill several platform capabilities: Interoperability, Security & Privacy, Developer Support, Data Management, Device Management and Services Management. Several authors focus on specific platform capabilities, such as security (Hytönen et al., 2022), interoperability (Marheine, 2021; Zyrianoff et al., 2021), or data management (Botta et al., 2020; Konduru & Bharamagoudra, 2021). Li et al. (2016) propose a reference architecture for IoT. Despite the abundance of IoT platform research, there is a research gap addressing the specific technological and economic requirements for designing technical smart service system architectures.

### 2.2 Architectural Requirements of Smart Services

Smart services, particularly in the context of IoT, are a growing area of interest in academic literature. These services represent the advanced capability of IoT systems to offer not just connectivity but intelligent, context-aware interactions and processes (Herterich et al., 2023; Meierhofer et al., 2020). The literature explores various facets of smart service architecture utilizing IoT, from theoretical frameworks to practical applications and future research directions.

Smart services are advanced functionalities that leverage the IoT's connectivity and data-processing capabilities to provide personalized, context-aware services to users (Beverungen et al., 2019). These services go beyond basic functionality, integrating with diverse devices and data sources to offer seamless and intuitive user experiences.

The literature discusses several architectural requirements regarding the design and implementation of smart services. Key requirements include data security. Implementing algorithms for different security levels to ensure the integrity and privacy of the data exchanged within an IoT smart services environment is crucial (Jerald et al., 2017). A second requirement is scalability. For smart services to be effectively deployed, especially in smart cities, hierarchical modeling of cloud-based IoT services is proposed to manage service real-time data and subscriptions (Taherkordi & Eliassen, 2016). A third requirement is reliability. It must be ensured that smart services can operate offline in environments with weak signals, thus reducing manual operations and enhancing convenience across various smart scenarios (Wu & Huang, 2020).

Several architecture approaches have been proposed to meet the unique requirements of smart services in IoT contexts: Utilizing Service-Oriented Architecture (SOA) principles for IoT enables the creation of flexible and modular services, including the orchestration of specific services based on user location and needs (Patel, 2018). SOA, however, may lead to scalability and performance issues. Fog computing is a distributed computing paradigm that provides computing, storage, and networking services between end devices (such as IoT devices) and cloud data centers (Atlam et al., 2018). Kum et al. (2017) propose a fog computing framework that acts as a service gateway for IoT applications, addressing interoperability and the vertical silo issues of smart services.

Despite considerable advancements, the field of smart service architecture in IoT continues to present open research avenues. One such research challenge is the integration of artificial intelligence within IoT frameworks to enhance service intelligence and user experience (Dremel et al., 2017; Poniszewska-Maranda et al., 2019). Another open research challenge is security, for example the design of architectures for secure smarter services, emphasizing flexibility, configurability, and scalability in IoT scenarios (Cirani et al., 2014). In summary, the design of IoT architectures supporting the design and delivery of smart services remains an ongoing research challenge.

## 3     Methodology

We adopt a design science research approach to deduce design principles for technical smart service architectures (Hevner & Chatterjee, 2010; Peffers et al., 2007). Design principles aim to generate knowledge of how design artifacts such as systems architectures can and should be constructed or arranged (i.e., designed), usually by human agency, to achieve a desired set of goals. This design knowledge includes understanding how to structure and construct various systems, how to model processes, and how to align systems with strategies (Dellermann et al., 2019; Vom Brocke et al., 2020).

Our methodological approach includes several empirical data collection methods (interviews, workshops) in two large industrial product firms to deduce solution requirements and validate design decisions. The interviews were conducted with interdisciplinary experts, also including customers. Workshops were usually done according to a design thinking methodology, finding out the pains and gains of the customers (Osterwalder, 2004). In later phases of the process, the project team was split into technical- and business-oriented groups who could develop content and requirements separately and at their own pace. Regular checkpoints were established to synchronize the efforts.

Customer co-creation was important in two separate phases of the process (see fig 1). The customers' input was extremely valuable in the two early phases, when the problems were identified, and the objectives defined. Later in the process, when the design principles were demonstrated and evaluated, the smart service development required testing in an original industrial facility on real customer data. In Figure 1 we show the corresponding process phases.

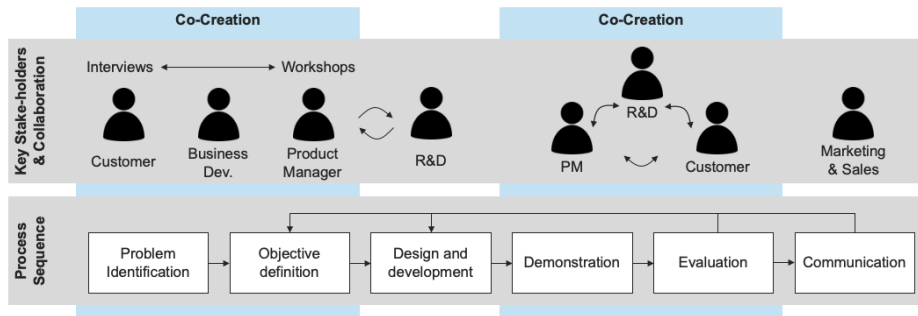

**Fig. 1.** Process chain as adapted from (Peffers et al., 2007). In the top-row we show the key stakeholders in each process step.

## 4    Results

Our primary goal is to uncover the design principles that shape corporate system architecture for companies entering the smart services market. We follow the process chain outlined by (Peffers et al., 2007), focusing on two firms active in high-precision measurement instrumentation. These companies traditionally manufacture and sell sensing devices that output data, forming the foundation for future smart services across various industries. Currently, their product portfolios are largely hardware-based, with accompanying software for data acquisition and basic processing. While some devices have basic cloud connectivity, most of the installed base and legacy products are not connected, relying instead on one-time sales with traditional pricing models. Service is provided through a large workforce of technicians who perform on-site interventions when needed.

These companies are at a moderate level of digitalization, with key processes like Enterprise Resource Planning (ERP), Customer Relationship Management (CRM) and Continuous Integration – Continuous Delivery (CI-CD) already digitized. However, sales, service processes, and R&D toolchains remain focused on traditional methods. Expanding the product portfolio to include smart services represents a logical next step in their digitalization and innovation journey, aligning with current market trends and creating a clear path for business growth.

This transformation presents significant challenges, particularly in the areas of strategy, processes, organization, spirit, and culture, as detailed by Frankenberger et al. (2020). We identify the technical corporate system architecture as a crucial enabler for bringing smart services to market.

### 4.1 Step 1: Identify Problem

Identifying the problem calls for an inter-disciplinary approach, involving co-creation with the customers. Since different stakeholders inside and outside the company have a very diverse level of knowledge about the technical realization of smart services, this process can be lengthy and involves a variety of different discussions and workshops, to establish a common ground about targets and challenges. We identified several problems to be solved.

**P1. Regulatory and compliance issues:** Highly regulated industries (e.g., food, pharma, trade, manufacturing) face significant requirements and potential costly re-certification for changes, especially with evolving artificial intelligence (AI) and data regulations like the EU Data Act and AI Act, necessitating future-proof service development.

**P2. Data privacy concerns:** In traditional industries, while customers are becoming more open to sharing data with smart service providers, concerns about data privacy and competitive advantage still hinder the use of customer data for AI-based service development.

**P3. Security concerns:** Connectivity is crucial for smart services but poses security challenges and demands frequent updates, especially in critical infrastructure where disruptions must be carefully managed.

**P4. Lack of standardization:** The core of smart service architecture lies in flexible modules, interfaces, and data pipelines that can integrate with existing systems and adapt to future market changes.

**P5. Integration with legacy systems:** Integrating existing software solutions into a new smart services framework is challenging due to the evolution in programming methods over recent years.

**P6. One-fits-all inside the corporation:** Traditional large companies often struggle with siloed structures and diverse customer requirements, leading to inconsistent smart service capabilities and unsatisfactory one-size-fits-all solutions.

**P7. Integration with corporate systems, ERP, CRM, CI-CD etc.:** To avoid immense manual effort, smart services must be fully digitalized and automated within the corporate process chain for smooth value creation for both providers and customers.

**P8. Currently uncommon, new business model:** Smart services transition from one-time sales with cost-plus pricing to value-based pricing and recurring revenue models, posing challenges for ERP systems not designed for these new business models.

**P9. Cost-control in development:** Implementing smart services in corporate systems requires significant R&D investment in skilled personnel, technology, and software, necessitating convincing decision-makers of the solution's value and ensuring a reasonable timeline for continuous, monetizable output.

## 4.2    Step 2: Define Objectives

The objectives are requirements, which target at capturing certain gains, as identified with the initial workshops. The following objectives were identified:

**O1. Global operation and compliance:** In a global corporation, ideally, every service should be sellable worldwide. However, the global smart services market is fragmented by diverse regulations and tools. Eastern and western markets align on different regulatory frameworks. Additionally, compliance with trade laws, technology transfer rules, and US export regulations is necessary. Cloud service preferences also vary, with globally spread customers favoring different providers. Therefore, a corporate architecture must remain flexible to deploy solutions across both various cloud platforms.

**O2. Digitalization and automation:** The value of any data-based service or solution is maximized, if it does not only help to create additional revenue for the company, but also helps to realize cost savings. Complete digitalization and automation are a great chance in this respect, because it helps to decrease effort in operations. Ideally no person is needed in regular operations and a solid CI/CD pipeline is ready for developments, updates, deployment and customer service.

**O3. Seamless flow of data and money:** When cloud services are involved, every data flow incurs costs, as pricing is typically based on data volume or operation time. Therefore, data pipelines must be optimized to align with the cost structure. It is crucial to monitor data traffic closely and design file formats efficiently. Computation resources across the technology stack should be used to pre-process and aggregate data, minimizing costs at cloud interfaces.

**O4. Serviceability and Maintainability:** Smart services, typically powered by algorithms and AI, offer the advantage of remote maintenance by a globally distributed workforce. Secured interfaces to customer systems are essential for this, along with self-diagnosis features for quick service interventions. Providers must establish a new service organization skilled in remote support and maintaining algorithms and models.

**O5. Scalability and modularization:** Any modern architecture is built up in a modular way, so that the single modules can be exchanged and developed in the respective speed, which is desired for their application. Inside the company, it makes sense to check for synergies across siloed structures, divisions and business units. Different parts of the same company might be at different levels of cloud readiness and digitalization.

**O6. Salability and monetizability:** A smart service is a digital product that requires a digital sales channel. Ideally, it should be available on a platform like a webshop, allowing customers to order, download, and deploy it without needing a sales representative. This process should be fully automated and integrated with ERP and CRM systems for seamless billing and monetization. Additionally, the pricing scheme must align with what the customer is willing and able to pay.

### 4.3    Step 3: List Design Principles

We identified a total of eight design principles discussed in the following.

**DP1. Allow deployment flexibility:** One should use a federated system and distributed processes. Containerized modules allow an agile deployment across different layers, depending on available computation resources. To maintain flexibility, one should avoid hard-coding to specific layers, allowing smart service components to be distributed throughout the stack. For instance, in an AI-based smart service for industrial production, data can be collected and aggregated at lower stack levels, ensuring that only non-critical information reaches the cloud, meeting customer and regulatory requirements. AI models can be split to run both below and above the data-barrier. Experience shows there is almost always a way to implement smart services within these constraints.

**DP2. Enable integration with customer system:** As providers develop smart services and system architectures, customers are doing the same. Large industrial facilities may already have IoT systems in place for monitoring production. To be convenient and cost-efficient, any smart service must integrate seamlessly with the customer's architecture. Data exchanges between provider and customer systems must use the correct file format and standardized interfaces.

**DP3. Leverage customer co-creation:** Co-creating a service with a customer ensures the solution meets their needs. It is a valuable opportunity for knowledge- and data sharing. Since data belongs to the customer, collaboration is crucial for smart service providers to develop effective solutions. Smart services often rely on AI, which requires a close feedback loop between customer and provider to adapt models for smooth operation. Technically, co-creation is facilitated by a protected environment, deployable flexibly in a container on the customer's or provider's cloud.

**DP4. Adopt cost-optimized migration path for legacy integration:** Some existing software solutions should be enhanced with smart services or cloud connectivity. Success hinges on standardized interfaces. While legacy code may not be cloud-ready, isolating certain components and standardizing interfaces and data formats can clarify inputs and outputs. When feasible, these components can be containerized for flexible deployment. However, this can sometimes result in large containers requiring significant computational resources, limiting flexibility. It is crucial to prioritize modernization efforts by focusing on "low-hanging fruits" and solutions that offer the most value to both the customer and provider.

**DP5. Enable scalability:** Initially, smart services handle small data volumes and few modules. However, as they succeed, system requirements can quickly scale, pushing technological limits. To ensure flexibility, cloud-deployment allows easy scaling up or down. Early consideration of data volume scaling is crucial for choosing suitable file formats and effective data aggregation methods.

**DP6. Embrace external solutions through modularity:** Whenever a solution is ready in the market for reasonable conditions (e.g. prize, lifetime, service, etc.) it should be preferred over developing an own solution. The biggest building block is the integration of an IoT platform.

**DP7. Build and use standardized interfaces:** Chose established, standard technologies. Agree on a set of data formats and interfaces inside the entire architecture and make sure that this is followed inside the company.

**DP8. Plan for seamless business model integration:** The business model may rely on various metrics like data volume, operation time, or event counts. Each data-based service must measure and log these metrics, feeding them through a message queue and application programming interface (API) to the ERP system for accurate billing. A consistent pipeline, from sensor to cloud, must integrate data transfers and trigger signals.

### 4.4   Step 4: Demonstration

To showcase the design principles, they were applied in a corporate smart service architecture project. The resulting architecture is shown in Figure 2.

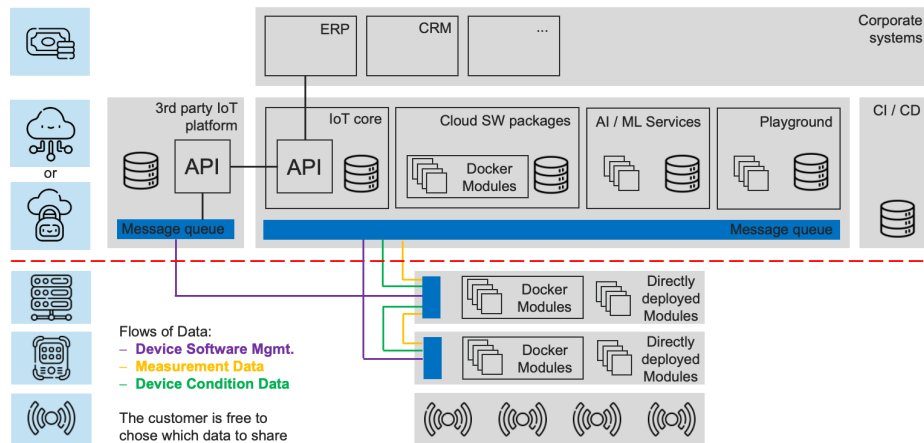

**Fig. 2.** Simplified representation of the corporate architecture along the layers of the technology stack. If software modules can be deployed in Docker containers, they can be deployed close to different computation resources. The customer can choose which kind of data they want to share.

In a corporate demo, a specific aspect of the architecture was implemented, showcasing it at corporate events and gathering interdisciplinary feedback. Furthermore, the architectural approach was used in several customer projects. Co-creation with customers provided valuable insights and feedback.

### 4.5 Step 5: Evaluation

To evaluate our design principles, we first check, if all the identified problems are solved, and if all the defined objectives are covered. The results are summarized in Table 1. Each of our design artifacts can deliver a solution to several of the problems and objectives.

| Problem (P) | Objective (O) | Artifact | Answering P or O |
|---|---|---|---|
| 1. Regulatory and compliance issues | 1. Global operation and compliance | 1. Deployment flexibility | P 1,2,3, O 1,5 |
| 2. Data privacy concerns | | 2. Integration with customer | P 1,2,3,5,7, O3,5,6,7 |
| 3. Security concerns | 2. Digitalization and automation | 3. Customer Co-Creation | P 7, 8, O 1, 2, 3, 4, 5, 6, 7 |
| 4. Lack of standardization | 3. Seamless flow of data and | 4. Legacy integration | P 5, 7, 9, O 2, 3, 4, 5 |
| 5. Integration with legacy systems | money | 5. Scalability | P 7, 8, 9, O 5 |
| 6. One-fits all inside the corporation | 4. Serviceability and | 6. Modularity | P 4, 5, 6, 9, O 4, 5 |
| 7. Integration with corporate systems | Maintainability | 7. Standardized interfaces | P 4, 5, 6, 7, 9, O 1, 2, 3, 5 |
| 8. New business model | 5. Scalability and modularization | 8. Seamless business model | P 7, 8, O 2, 3, 4, 6, 7 |
| 9. Cost-control in development | 6. Sellability and monetizability | integration | |

**Table 1.** Summary of problems, objectives and the resulting design artifacts.

## 5 Discussion

A research gap exists in designing IoT architectures for smart services. Based on two architecture projects, we introduce and validate our design principles that guide smart service architecture design, contributing valuable knowledge to this dynamic field. These principles offer crucial guidance for designing and customizing smart service architectures, helping providers create fit-for-purpose systems.

Regarding contributions to practice, the presented design principles guide the design of general-purpose smart service architectures that cater for heterogeneous smart service demands. In this sense, they make transparent and help to maneuver different design tradeoffs shown in Fig. 3. These tradeoffs can be categorized as technical, business related and practical.

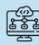

**Fig. 3.** Tradeoffs in a technology stack representation from the sensor to the cloud. On top, we show the data-based service layer, which is focused on monetization.

**Speed vs. data volume:** Running algorithms close to the sensor allows for low-latency. However, cloud data transfer introduces unavoidable latency. Conversely, sensor-level computation is limited by space, cost, and heat dissipation, so large data volumes must be processed higher up the technology stack.

**Full automation vs. scalability:** Low-latency processing near the sensor allows easy automation using control electronics. However, the number of devices that can be integrated is limited by available hardware interfaces, leading to high maintenance needs and scalability limits. In contrast, cloud services offer virtually unlimited scalability.

**Closed-loop control vs. analytics:** Low latencies near the sensor allow quick responses to system changes with closed-loop control. However, integrating predictive analytics, a key smart service, requires connection to higher technology stack levels for adequate data volume and computation power.

**Established business model vs. new business model:** For many providers, the traditional business focuses on hardware at the lower technology stack. The new focus on monetizing smart services at the top of the stack demands a complete shift in the business model.

**Transactional sales vs. regular payment schemes:** Traditional business models rely on one-time transactional sales, while smart services typically use recurring revenue and regular payment schemes.

**Running systems vs. investment needs:** Traditional product portfolios are well-established, while innovating toward smart services requires significant investment across various corporate areas and systems.

**One-time installations vs. regular updates:** Old stand-alone devices were never internet-connected, so they required no updates as long as they functioned properly. However, connecting devices to the internet or cloud introduces security risks, necessitating regular updates to maintain system security.

**Certified systems vs. re-certification:** In highly regulated industries, system changes, especially software, can require significant re-certification efforts. Smart services and connectivity add complexity, requiring compliance with additional regulations.

**Established processes vs. ramp-up of new organizations:** Introducing smart services impacts not just the software architecture but also the entire corporate process landscape and toolchain. New talent and teams are required across multiple areas of the company.

## 6    Conclusion

Despite its contributions to theory and practice our work does not come without limitations. First, the development and validation of the design principles was conducted on the empirical basis of two companies. This allowed us deep insights into stakeholder interests, contributions and perceptions throughout design, implementation and validation. Future research is required to evaluate the developed design principles on a broader empirical basis. Second, our validation was limited to selective proof of concepts. Further research is required to study the effects of architecture design under the proposed design principles for longer time periods and broader applications. Particularly relevant will be the robustness regarding future technology dynamics. Third, our research was limited to applications in the manufacturing industry. Smart service architectures are also used in other industries such as construction or utilities. Further research is required to test and validate the generalizability of our findings.